\documentclass[twocolumn]{aastex631}

\newcommand{\ketju}{\textsc{Ketju}}                           
\newcommand{\gadget}{\textsc{gadget-4}}                       

\newcommand{\Msun}{\ensuremath{\mathrm{M}_{\sun}}}            
\newcommand{\kmps}{\ensuremath{\mathrm{km}\,\mathrm{s}^{-1}} }
\newcommand{\Reff}{\ensuremath{R_\mathrm{e}}}                 
\newcommand{\rapo}{\ensuremath{r_\mathrm{apo}}}               
\newcommand{\rapop}{\ensuremath{r_\mathrm{apo,proj}}}         
\newcommand{\vk}{\ensuremath{v_\mathrm{kick}}}                
\newcommand{\sigcore}{\ensuremath{\sigma_{\star,0}}}          
\newcommand{\rdetect}{\ensuremath{r_\mathrm{d}(\vk)}}         

\newcommand{\mbound}{\ensuremath{M_\mathrm{BRC}}}           
\newcommand{\BRCR}{\ensuremath{R_\mathrm{e,BRC}}}         
\newcommand{\BRCV}{\ensuremath{\sigma_{\star,\mathrm{BRC}}}}    

\newcommand{\dd}[1]{\ensuremath{\mathrm{d}#1}}                
\newcommand{\dv}[2]{\ensuremath{\frac{\dd{#1}}{\dd{#2}}}}     
\newcommand{\dnv}[3]{\ensuremath{\frac{\mathrm{d}^#1#2}{\dd{#3}^#1}}}  
\newcommand{\ordinal}[1]{\ensuremath{#1^\mathrm{th}}}         

\usepackage{xcolor}

\usepackage{amsmath}
\usepackage{mathrsfs}

\submitjournal{ApJ}

\graphicspath{{./}{figures/}}

\begin{document}

\title{Caught in the act: detections of recoiling supermassive black holes from simulations}

\correspondingauthor{Alexander Rawlings}
\email{alexander.rawlings@helsinki.fi}

\author[0000-0003-1807-6321]{Alexander Rawlings}
\affiliation{Department of Physics,
Gustaf H\"allstr\"omin katu 2, FI-00014, University of Helsinki, Finland}

\author[0000-0001-8741-8263]{Peter H. Johansson}
\affiliation{Department of Physics,
Gustaf H\"allstr\"omin katu 2, FI-00014, University of Helsinki, Finland}

\author[0000-0002-7314-2558]{Thorsten Naab}
\affiliation{Max-Planck-Institut f\"ur Astrophysik, Karl-Schwarzchild-Str 1, D-85748 Garching, Germany}

\author[0000-0001-8789-2571]{Antti Rantala}
\affiliation{Max-Planck-Institut f\"ur Astrophysik, Karl-Schwarzchild-Str 1, D-85748 Garching, Germany}

\author[0000-0003-2868-9244]{Jens Thomas}
\affiliation{Max-Planck-Institut f\"ur extraterrestrische Physik, Giessenbachstrasse, D-85748 Garching, Germany}
\affiliation{Universit\"ats-Sternwarte M\"unchen, Scheinerstrasse 1, D-81679 M\"unchen, Germany}

\author[0000-0001-6564-9693]{Bianca Neureiter}
\affiliation{Max-Planck-Institut f\"ur extraterrestrische Physik, Giessenbachstrasse, D-85748 Garching, Germany}
\affiliation{Universit\"ats-Sternwarte M\"unchen, Scheinerstrasse 1, D-81679 M\"unchen, Germany}



\begin{abstract}
We study the detectability of supermassive black holes (SMBHs) with masses of $M_{\bullet}\gtrsim 10^{9}\,\Msun$ displaced by gravitational wave recoil kicks $(v_{\rm kick}=0\mathrm{-}2000\,\kmps)$ in simulations of merging massive $(M_{\star}>10^{11}\,\Msun)$ early-type galaxies. The used \ketju{} code combines the \gadget{} fast multiple gravity solver with accurate regularised integration and post-Newtonian corrections (up to PN3.5) around SMBHs. The ejected SMBHs carry clusters of bound stellar material (black hole recoil clusters, BRCs) with masses in the range of 
$10^6 \lesssim M_{\text{BRC}} \lesssim 10^7\,\Msun$ and sizes of several $10\,\mathrm{pc}$.
For recoil velocities up to 60\% of the galaxy escape velocity, the BRCs are detectable in mock photometric images at a Euclid-like resolution up to redshift $z \sim 1.0$. 
By Monte Carlo sampling the observability for different recoil directions and magnitudes, we predict that in $\sim20\%$ of instances the BRCs are photometrically detectable, most likely for kicks with SMBH apocentres less than the galaxy effective radius. BRCs occupy distinct regions in the stellar mass/velocity dispersion vs. size relations of known star clusters and galaxies. 
An enhanced velocity dispersion in excess of $\sigma \sim 600\,\kmps$ coinciding with the SMBH position provides the best evidence for an SMBH-hosting stellar system, effectively distinguishing BRCs from other faint stellar systems.
 BRCs are promising candidates to observe the aftermath of the yet-undetected mergers of the most massive SMBHs and we estimate that up to 8000 BRCs might be observable below $z\lesssim 0.6$ with large-scale photometric surveys such as Euclid and upcoming high-resolution imaging and spectroscopy with the Extremely Large Telescope. 

\end{abstract}



\section{Introduction} \label{sec:intro}

Both observations \citep[e.g.][]{tran2005,trujillo2006,mcintosh2008,vandokkum2010,vanderwel2014}, and theoretical models \citep[e.g.][]{delucia2006,Naab2009,Johansson2012,laporte2013,behroozi2013,Rodriguez-Gomez2016,moster2018,moster2020} have long indicated that, in particular for massive early-type galaxies (ETGs) in dense environments, mergers of ETGs are frequent and dominate the mass assembly and size growth below redshifts of $z \lesssim 2$. Central supermassive black holes (SMBHs) are believed to reside 
at the centers of all massive galaxies \citep[e.g.][]{kormendy2013}, and if the host galaxies merge, subsequent sinking and coalescence of the SMBHs is expected to occur through a three-stage process \citep{begelman1980}. 

Firstly, dynamical friction \citep{chandrasekhar1943} brings the SMBHs down from kiloparsec scales to parsec-scale separations. In the second phase, the SMBH separation is further reduced through sequential three-body interactions with the surrounding stellar component \citep{hills1980,quinlan1996}. 
In the final phase, at subparsec separations, gravitational wave (GW) emission becomes the dominant mechanism for the loss of the remaining orbital energy and angular momentum, thus driving the SMBHs to coalescence \citep{peters1963}. 
The sinking and resulting mergers of SMBHs provide plausible explanations for the observed low scatter in the SMBH scaling relations \citep[e.g.][]{hirschmann2010,jahnke2011}, the formation of the observed faint, diffuse central cores with missing starlight found at the centers of massive ETGs \citep[e.g.][]{milosavljevic2001,hoffman2007,rantala2018,nasim2021b}, and repeated core scouring can also increase the effective radii of massive ETGs by up to a factor of two \citep[e.g.][]{rantala2019,rantala2024}. 

Non-isotropic GW emission from the SMBH binary carries linear momentum away from the system \citep[e.g.][]{bekenstein1973,gonzalez2007}, which at the time of coalescence, imparts a recoil (or kick) velocity to the remnant SMBH, varying from tens to thousands of kilometers per second in magnitude \citep{campanelli2007,gonzalez2007b}.
The magnitude of the recoil velocity depends on the progenitor SMBH mass difference and the relative difference between the SMBH spin vectors, with largest kicks occuring for equal-mass SMBHs with anti-aligned spins \citep{gonzalez2007b}.
As the central escape velocity of a typical massive ETG is $v_{\rm esc}\sim 2000 \,\kmps$, a non-negligible fraction of galaxies might have ejected their SMBHs \citep[e.g.][]{madau2004,mannerkoski2022}, however not so many as to introduce considerable scatter in the observed SMBH scaling relations \citep{volonteri2007}. Numerical simulations have also shown that the gravitational recoil-driven ejection of the coalesced SMBH can further increase the size of the galaxy core beyond that of the initially binary-scoured core \citep[e.g.][]{gualandris2008,nasim2021b,khonji2024,rawlings2025}.

Recent results from pulsar timing array (PTA) observations show tentative evidence for a stochastic gravitational wave background signal consistent with the merging of massive SMBHs $(M_{\bullet}\gtrsim 10^{8} M_{\odot})$ \citep[]{agazie2023,2023Xu,zic2023,antoniadis2024}. 

Presently, there is no direct observational confirmation for SMBH mergers (unlike stellar mass BHs) and resulting recoil kicks.
However, several observational studies have hinted at the existence of recoiling SMBHs, generally through an offset from the center of the galaxy in either position, in velocity, or both \citep[e.g.][]{komossa2008,Eracleous2012,comerford2014,pesce2021,hogg2021,Chiaberge2025}. 
Attributing an escaping SMBH to GW recoil, as opposed to a three-body interaction between a binary SMBH and a third SMBH, is not always straightforward, as evidenced in the recent detection by \citet{vandokkum2023} of a runaway SMBH candidate identified by its interaction with the surrounding circumgalactic medium. Gravitationally recoiling SMBHs can also affect their accretion rates and thus quasar lifetimes, as the SMBHs are offset from the central gas density peak, potentially also inducing additional scatter in the observed SMBH scaling relations \citep[e.g.][]{blecha2008,blecha2011}.

Using numerical simulations, \citet{merritt2009} proposed that a recoiling SMBH would carry a retinue of bound stars with it \citep[see also][]{boylan2004,oleary2009}. 
These stars would form a stellar system around the recoiling SMBH, which we will term a `black hole recoil cluster' (BRC) in this work, with characteristic sizes and luminosities similar to globular clusters.
However, in extreme cases, when kicked from massive galaxies, the BRC masses could even approach that of ultracompact dwarf galaxies. 
The defining characteristic of BRCs surrounding SMBHs and what sets them apart from other compact stellar systems is their very high internal stellar velocity dispersions, which scales with recoil velocity, and is typically two orders of magnitude higher than other compact stellar systems. Observational studies have already attempted to identify such massive clusters in massive cored ETGs \citep{Burke-Spolaor2017,Gultekin2021}. In addition, searches for lower-mass compact stellar systems have also been conducted in the Milky Way \citep[e.g.][]{Greene2021,Wu2024}.

In this paper, we utilise self-consistent gas-free simulations of merging ETGs with recoiling SMBHs that are run using the GADGET-4 based regularised tree code \ketju{} \citep{rantala2017,rantala2020,mannerkoski2023}. Specifically, we use the large simulation sample presented in \citet{rawlings2025} in which the SMBH kick velocity was systematically varied to study its impact on the host galaxy. 
Earlier, \ketju{} has been used to study the formation of cores in gas-free simulations, but without the contribution of SMBH recoils \citep{rantala2018,rantala2019,rantala2024,mannerkoski2023,partmann2024}. 

The aims of this paper are as follows. Firstly, we study the properties of the black hole recoil clusters (BRCs) that are surrounding the recoiling SMBHs. Secondly, using a synthesis of photometric and kinematic observations, we study the potential detectability of BRCs in large-scale extragalactic surveys, and the impact on the host galaxy as well.
Finally, by combining photometric survey data and more detailed spectral kinematic data, we present a workflow that could help identify the unique observational signatures of BRC. 

This paper is structured as follows. In Section \ref{sec:methods}, we present the numerical simulations used in this investigation. 
We construct mock photometric observations of the BRCs in Section \ref{sec:photo_obs} and analyze their detectability both for BRC kicked orthogonally and off-axis with respect to the line-of-sight. In Section \ref{sec:kinematics} we construct mock kinematic maps and discuss observable features that could reveal the presence of recoiling SMBHs and BRCs. Finally, we discuss our results and present our conclusions in Section \ref{sec:discussion}.

\section{Numerical Methods} \label{sec:methods}
\subsection{Simulation code}\label{ssec:sim_code}
The simulations presented in this study were run with the new public version of the \ketju{}\footnote{\url{https://www.mv.helsinki.fi/home/phjohans/group-website/research/ketju/}} code \citep{rantala2017,rantala2020,mannerkoski2023}, which is coupled to the GADGET-4 code \citep{springel2021}. 
\ketju{} adds a small algorithmically regularised region (i.e., the \ketju{} region) around each SMBH, in which the dynamics of SMBH and the surrounding stellar component is integrated using the MSTAR integrator \citep{rantala2020}. The dynamics of stellar particles beyond this small region and the dynamics of all dark matter particles are then computed with the GADGET-4 fast multipole method (FMM) gravity solver with second-order multipoles. We also employ hierarchical time integration, which allows for mutually symmetric interactions and manifest momentum conservation. \ketju{} also includes post-Newtonian (PN) correction terms up to order PN3.5 with optional spin terms between each pair of SMBHs \citep{blanchet2014} and the fitting formula of \citet{zlochower2015} for recoil kick velocities.     

\subsection{Merger simulations}\label{ssec:merger_sims}
We use the gas-free isolated major merger simulations of massive ETGs from \citet{rawlings2025} and summarise here the main properties of the simulation setup; for full simulation details, please refer to \citet{rawlings2025}.

Each progenitor galaxy is represented by an isotropic multicomponent system and consists of a stellar component with a total mass of $M_\star\sim 1.38\times10^{11}\,\Msun$ embedded in a dark matter (DM) halo with a total mass of $M_\mathrm{DM} = 2.5\times10^{13}\,\Msun$ \citep{moster2010}. At the center of the halo, we place an SMBH with a mass of $M_{\bullet,0} = 2.93 \times 10^9\,\Msun$ \citep{sahu2019} and zero initial velocity. Both the stellar and DM components each follow a \citet{hernquist1990} profile, with a scale radius of $a_{\star}=3.9 \ \rm kpc$ and  $a_{\star}=245 \ \rm kpc$, respectively. The density profile $\rho_{i}$ of each component is then given by:
\begin{equation}\label{eq:hernquist}
   \rho_{i}(r)=\frac{M_{i}}{2\pi}\frac{a_{i}}{r(r+a_{i})^{3}}.
\end{equation}
The stellar and DM particle masses are $m_\star=5.0\times10^4\,\Msun$ and $m_\mathrm{DM} = 5.0\times10^6\,\Msun$, respectively.
The simulations are performed with \ketju{}, where SMBH-star and SMBH-SMBH interactions are non-softened, and all other interactions are softened: $\varepsilon_\star=2.5\,\mathrm{pc}$ for stellar particles, and $\varepsilon_\mathrm{DM}=200\,\mathrm{pc}$ for DM particles.
All stellar particles beyond $3\varepsilon_\star=7.5\,\mathrm{pc}$ of an SMBH, and all DM particles, are integrated using \gadget{} \citep{springel2021}.

The simulations proceed in two stages.
First, the galaxies merge on a nearly-radial orbit from an initial separation of $30\,\mathrm{kpc}$ with impact parameter of $2\,\mathrm{kpc}$.
During the merger the SMBH binary scours a core of radius $r_{\mathrm{b},0}=0.58\,\mathrm{kpc}$, until the SMBH binary coalesces at a time $t_\mathrm{coal}$.
The stellar velocity dispersion within the core radius $r_{\mathrm{b},0}$ at $t_\mathrm{coal}$ is $\sigcore=270\,\kmps$.
Next, 31 independent `child' copies of the simulation at $t_\mathrm{coal}$ are created, where the coalesced SMBH ($M_\bullet=2M_{\bullet,0}=5.86\times10^9\,\Msun$) in each copy has a uniquely-prescribed recoil velocity between $0\,\kmps$ and $1800\,\kmps$, sampled in $60\,\kmps$ increments, arbitrarily along the $x$-axis.
The upper limit of the kick was chosen to match the escape velocity of the center of the merger remnant, $v_{\rm esc}=1800 \ \rm km \ s^{-1}$.

The simulation resumes until the SMBH has settled back to the centre of the galaxy or alternatively until the maximum simulation time of 3 Gyr has been reached. 

Whilst we neglect gas in these simulations, \ketju{} has previously been used to study SMBH merger timescales in hydrodynamical simulations, with the results from elliptical-elliptical mergers showing that the SMBH merger timescales and central stellar densities are similar for no-gas simulations and simulations that include both gas physics and models for stellar and BH feedback \citep{liao2024a,liao2024b}. 
As a consequence, we do not expect gas physics to play a major role in the dynamics of the recoiling SMBHs in the systems we study.

\subsection{Defining a black hole recoil cluster}
\begin{figure}[t]
    \centering
    \includegraphics[width=0.5\textwidth]{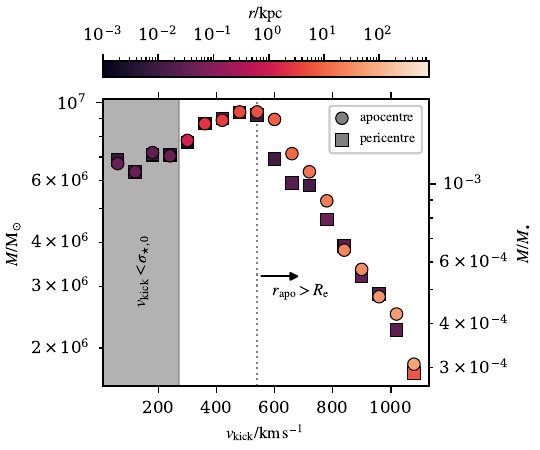}
    \caption{
        The intrinsic BRC mass $\mbound$ at apocentre (circles) and pericentre (squares) as a function of kick velocity.
        Points are coloured by the radial displacement of the SMBH from the galaxy centre.
        For $\vk \gtrsim 540\,\kmps$, $\mbound$ decreases exponentially.
        The grey-shaded region indicates recoil velocities less than the velocity dispersion of the SMBH binary-scoured core $\sigcore$.
    }
    \label{fig:bound}
\end{figure}
At the beginning of each child simulation, the kicked SMBH departs the centre of the galaxy, taking with it an entourage of gravitationally-bound stellar particles that constitute the black hole recoil cluster.

We begin by identifying those stellar particles within the SMBH influence radius $r_\mathrm{infl}$, determined as:
\begin{equation}
    M_\star(r < r_\mathrm{infl}) = 2 M_\bullet.
\end{equation}
We then require that the members of the BRC are strongly-bound\footnote{Note that to be bound, but not necessarily strongly, the particle binding energy must be greater than 0.} to the SMBH, such that their binding energy $E_\mathrm{b}$ is greater than the ambient stellar velocity dispersion $\sigma_\mathrm{amb}$:
\begin{equation}\label{eq:bound}
    E_\mathrm{b} \equiv \frac{GM_\bullet}{r_i} - \frac{1}{2} v_i^2 > \sigma_\mathrm{amb}^2
\end{equation}
for each stellar particle $i$ within the influence radius of the SMBH, where $v_i$ is the particle velocity relative to the SMBH, and $r_i$ is the displacement of the particle from the SMBH.
The ambient stellar velocity dispersion is determined within an aperture of $5r_\mathrm{infl}$ about the SMBH as:
\begin{equation}
    \sigma_\mathrm{amb} = \sqrt{\sigma_x^2 + \sigma_y^2 + \sigma_z^2},
\end{equation}
and is dependent on the radial distance of the SMBH from the galaxy centre.
The mass in stellar particles which satisfy \autoref{eq:bound} we denote as $\mbound$.

Strongly-bound particles at the first SMBH apocentre and following pericentre are shown in \autoref{fig:bound} as circle and square markers, respectively.
There is negligible difference in $\mbound$ at these two moments, indicating that these strongly-bound particles are travelling with the SMBH during its orbital excursion.
At most, $\mbound\simeq 2\times10^{-3}\,M_\bullet$.
For recoils $\vk\gtrsim 540\,\kmps$, $\mbound$ decreases with $\vk$, as the higher recoil velocity requires higher binding energies for the stellar particles to remain bound to the SMBH \citep[see also][]{merritt2009}.
Conversely for $\vk\lesssim 540\,\kmps$, $\mbound$ rises slightly.
This occurs due to the relative velocity term $v_i$ in \autoref{eq:bound}: for low $\vk$, whilst there are many particles bound to the SMBH with $E_\mathrm{b}>0$, they do not necessarily have $E_\mathrm{b}>\sigma_\mathrm{amb}^2$, and are hence not strongly-bound to the SMBH.
As $\vk$ increases above $\sigma_\mathrm{amb}$, a greater fraction of stellar particles bound to the SMBH typically also have  velocities above the ambient velocity dispersion.
Until $\vk\simeq 540\,\kmps \simeq 2\sigcore$ the recoil velocity of the SMBH is not so great so as to unbind a large amount of stellar mass from the recoiling SMBH (due to the kinetic energy of stellar particles increasing as $v_i^2$ from \autoref{eq:bound}), leading to the peak in $\mbound$ before declining for $\vk\gtrsim 2\sigcore$.

A large $\mbound$ is of interest when assessing the detectability of a recoiling SMBH with a non-zero orthogonal component of its velocity vector to the LOS \citep[e.g.][]{merritt2009,oleary2009} as the corresponding stellar luminosity may be detectable in photometric observations.

\section{Mock photometric observations}\label{sec:photo_obs}

\begin{figure*}
    \centering
    \includegraphics[width=\textwidth]{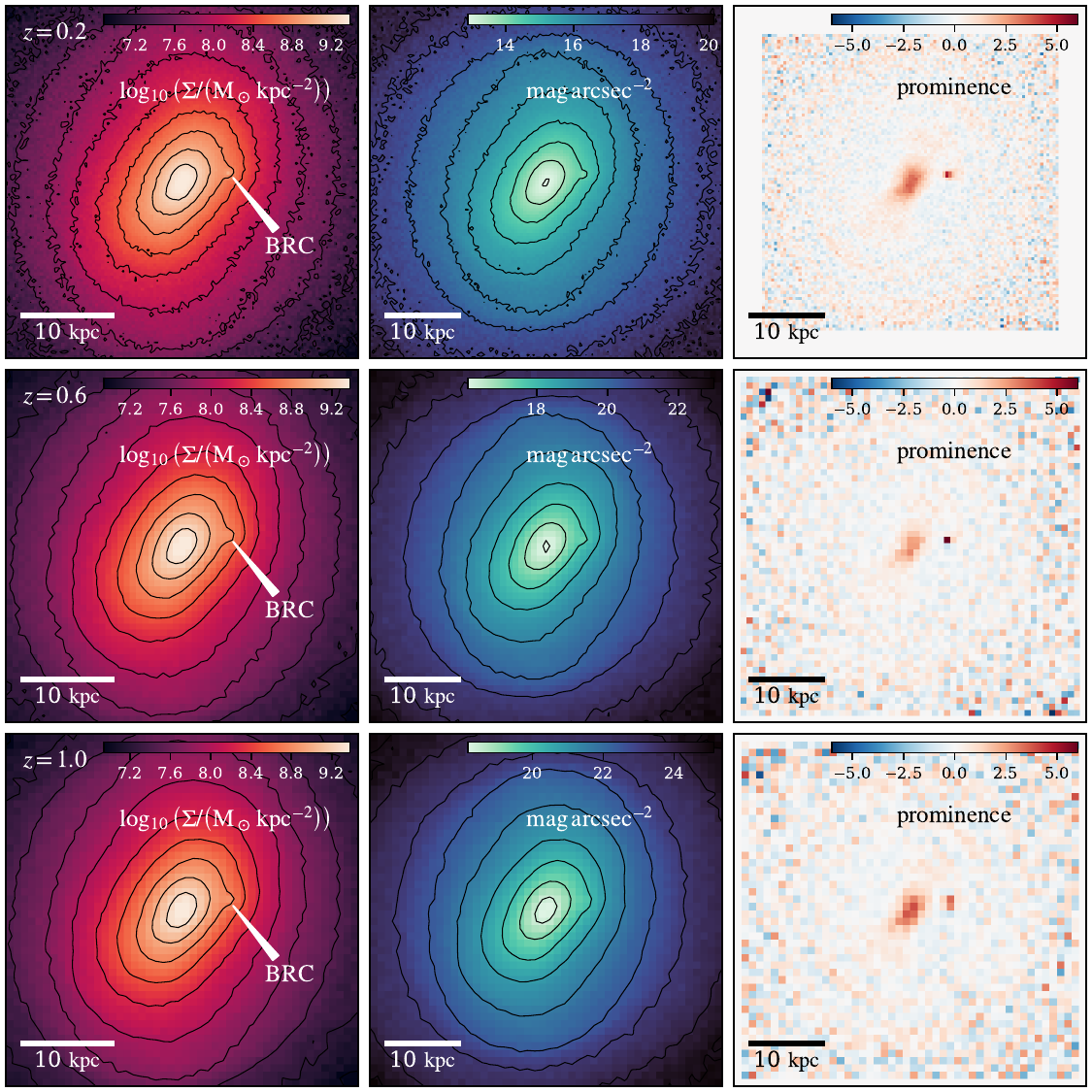}
    \caption{
        Black hole recoil cluster mock detections assuming different observation redshifts ($z=\{0.2,0.6,1.0\}$, each row respectively).
        \textit{Left column}: projected stellar mass density at a time $t\simeq0.03\,\mathrm{Gyr}$ after $t_\mathrm{coal}$, with the remnant SMBH and BRC moving along the positive $x$-axis with velocity $\vk=540\,\kmps$.
        \textit{Centre column}: a mock image of the remnant galaxy, in Euclid VIS band apparent magnitude.
        \textit{Right column}: the prominence from the apparent magnitude mock image.
        For each redshift, the BRC is seen to have a prominence of $\hat{K}\gtrsim 4.0$, and is thus detectable.
    }
    \label{fig:projdens}
\end{figure*}

\subsection{Creating mock photometric images}\label{ssec:mock_obs}
To ascertain if a peak in surface mass density would be detectable by instruments optimised for large-scale galaxy surveys, we create mock images for each simulated recoil velocity, specifically for each snapshot prior to the SMBH reaching its first apocentre.

We begin by calculating the luminosity of the stellar particles, using stellar population synthesis results from the code Binary Population and Spectral Synthesis \citep[BPASS,][]{stanway2018}, and assuming a \citet{chabrier2003} initial mass function with mass limits $0.1\,\Msun$ to $100\,\Msun$.
The age and metallicity of stellar particles is set to the median of the respective value of massive quiescent galaxies in the Large Early Galaxy Astrophysics Census (LEGA-C) presented in \citet{bevacqua2024}, namely $t_\mathrm{age}=3.645\,\mathrm{Gyr}$ and $Z=0.034$.

We create mock images of the galaxy merger remnant at three redshifts: a fiducal redshift of $z=0.6$, and one lower and one higher redshift: $z=0.2$ and $z=1.0$, respectively.
The redshift dictates the number of pixels used in the image construction, assuming a Euclid-like spatial sampling of $0\farcs101$/pixel.
The observation waveband corresponds to the visible of the Euclid VIS instrument, $550\,\mathrm{nm} \mathrm{-} 900\,\mathrm{nm}$ \citep{cropper2024}. 
The flux at each pixel (with the inclusion of cosmological dimming scaling as $(1+z)^{-4}$) is convolved with the Euclid transmission curves to obtain an apparent magnitude with $K$-correction.
Example mock images are shown in the centre column of \autoref{fig:projdens} for the $\vk=540\,\kmps$ case, where notably the BRC is still visible.

To quantify how distinct the BRC is from the background galaxy, we calculate a non-parametric prominence measure to identify a spatially-localised intensity increase in surface brightness associated with the BRC.
To do so, we use a low-pass Gaussian filter to remove high-frequency components (such as noise) from the image whilst permitting low-frequency components of the image \citep[e.g.][see \autoref{sec:app_prom}]{davies90}. 
An example of the prominence map is shown in the right panel of \autoref{fig:projdens}.
For the BRC to be detectable, we demand that the pixel hosting the BRC has a prominence $\hat{K}\geq4.0$ (see \autoref{sec:app_prom}), translating to the pixel having such a value only once in every $10^5$ instances by chance alone.

We also tested a parametric model-fitting technique, where the galaxy flux is modelled with a 2D S\'ersic profile and subtracted from the data, but found it to give inferior results to the prominence technique due to poor fitting in the outer regions of the galaxy.

In our experiment, the observability of the recoiling SMBH is dictated by two competing factors: obtaining a large enough projected distance to offset it from the isophotal centre of the merger remnant (correlating positively with $\vk$), and the BRC being massive and thus luminous enough to stand out in surface brightness (correlating negatively with $\vk$).
Almost certainly, the SMBH will recoil with a velocity that has a non-zero component parallel to the LOS,  $v_\parallel \neq 0$, reducing the apocentre distance of the BRC in projection for a given $\vk$.
The effect of off-axis motion we explore next.

\subsection{Off-axis recoil and detectable fraction}\label{ssec:offaxis}
\begin{figure*}[t]
    \centering
    \includegraphics[width=0.9\textwidth]{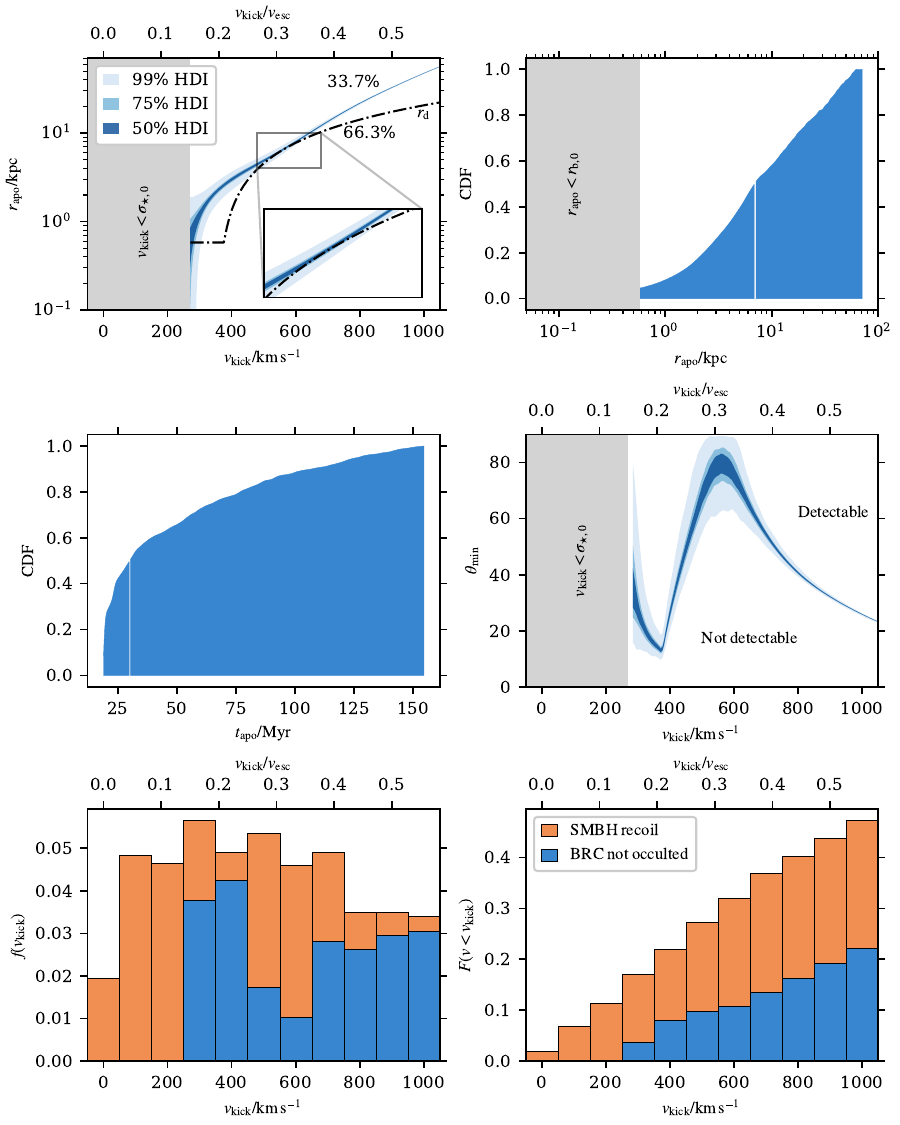}
    \caption{
        \textit{Top left}: Modelled relation between $\vk$ (for recoil $\geq \sigcore$) and apocentre, with blue regions indicating Bayesian highest density intervals (HDIs), and the dash-dotted line the detection threshold $\rdetect$, of which 33\% samples exceed.
        \textit{Top right}: Marginal cumulative distribution of apocentre distances.
        The CDF is segemented at the median.
        \textit{Centre left}: Transform sampled cumulative distribution of time to apocentre, with the CDF segmented at the median.
        \textit{Centre right}: Minimum angular offset from the LOS axis $\theta_\mathrm{min}$ that allows $\rapop$ to exceed $\rdetect$, colored by HDI.
        The space above the blue contours (including the region around $\vk\sim 350\,\kmps$) indicates angular offsets for which the BRC would be detectable.
        \textit{Bottom left}: Recoil velocity distribution binned in $100\,\kmps$ bins (orange), and weighted by the corresponding posterior draw of observational probability $\cos\theta_\mathrm{min}$ (blue)
        The $y$-axis depicts the frequency of a particular bin relative to the total number of posterior draws.
        \textit{Bottom right}: Corresponding cumulative distribution to the bottom left panel.
    }
    \label{fig:apos}
\end{figure*}
To determine the amount of off-axis motion the recoiling SMBH may have and still be detected, we take from each simulation the first apocentre for each kick velocity, for all simulations where the SMBH velocity vector changes signs at least once within the simulation time of $\sim3\,\mathrm{Gyr}$.
The detectability of the BRC is determined for each snapshot for each $\vk$ using the prominence method in \autoref{ssec:mock_obs}. 
For recoil velocities $\vk < \sigcore$ the SMBH does not travel beyond the stellar core radius of the galaxy, and is undetectable.
Similarly, for recoil velocities $\vk > 1080\,\kmps \simeq 0.6v_\mathrm{esc}$, the mass of the BRC is not enough to provide a detectable signature, in agreement with \citet{gualandris2008}.
Hence, we find that the minimum  distance for BRC detectability in kpc, $\rdetect$, depends linearly on $\vk$ as:
\begin{equation}
    \frac{\rdetect}{\mathrm{kpc}} = 57.8 \left(\frac{\vk}{v_\mathrm{esc}}\right) - 11.5
\end{equation}
for $\sigcore < \vk \lesssim 0.6v_\mathrm{esc}$: the BRC would not be detectable in large-scale surveys otherwise.

We fit the $\vk$--$\rapo$ relation using a Gaussian process (GP) regressor with \textsc{Stan} \citep{standevelopmentteam2018} in the top left panel of \autoref{fig:apos}.
To obtain a distribution of apocentres, we use in post-processing the \citet{zlochower2015} model for recoil velocity that takes as input the masses, spins, and phase-space coordinates of the SMBHs prior to coalescence.
We assume SMBH spins that are preferentially aligned with the SMBH binary angular momentum vector prior to merger by some angle $\phi$ \citep[which follows a right-skewed distribution with maximum at $\phi\sim25\fdg0$,][]{lousto2010}, and have a magnitude $\alpha_\bullet$ distribution given by
\begin{equation}
    P(\alpha_\bullet) \propto (1-\alpha_\bullet)^{4.686884-1} \alpha_\bullet^{10.5868-1},
\end{equation}
which is derived from repeated SMBH binary merger simulations that follow the spin evolution of the SMBHs \citep{lousto2010}.
We then perform transformation sampling of the obtained $\vk$ distribution using the GP regressor to obtain a distribution of apocentre distances, from which we find that $\rapo$ exceeds $\rdetect$ with a detectable BRC in 33\% of instances.
This we take as an upper limit to the number of galaxy merger remnants with an SMBH ejected to beyond $\rdetect$: an unrealistically optimistic case where each SMBH is ejected to an apocentre orthogonal to the LOS axis.
The cumulative marginal distribution of apocentres is shown in the top right panel of \autoref{fig:apos}: typically, the SMBH has a maximum apocentre of $\sim7\,\mathrm{kpc}$.
The typical time to reach apocentre is $\sim30\,\mathrm{Myr}$ following SMBH binary coalescence.

We next perform a second transformation sampling to construct the distribution of the minimum angular offset $\theta_\mathrm{min}$ from the apocentre distribution which results in the SMBH obtaining a projected apocentre distance\footnote{Here we assume that the observability is equal if $\theta_\mathrm{min}$ is directed towards or away from the observer. In an actual galaxy with dust obscuration, the observability would most likely be higher if the SMBH was recoiling towards the observer rather than away.} of $\rapop(v) \geq \rdetect$, shown in the centre right panel of \autoref{fig:apos}.
We define $\theta_\mathrm{min}$ relative to the LOS axis:
\begin{equation}
    \theta_\mathrm{min} = \sin^{-1}\left( \frac{\rdetect}{\rapo} \right).
\end{equation}
By randomly-sampling angles $\tilde{\theta}$ from the LOS axis, we find that in $22\%$ of instances, the projected apocentre distance of the recoiling SMBH exceeds $\rdetect$.

The dependence of $ \theta_\mathrm{min}$ on $\vk$ has a distinct double-dip feature and characteristic rise at $\vk\simeq 540\,\kmps$.
The first dip can be explained by considering the BRC mass $\mbound$ from \autoref{fig:bound}: for $\vk\lesssim 540\kmps$, the BRC mass increases, reaching a maximum at $\vk\simeq540\,\kmps$.
As the recoil velocity increases however, the SMBH is displaced to continuously larger $\rapo$, thus the angular offset required to achieve $\rapop \geq \rdetect$ decreases.
For $\vk\simeq 540\,\kmps$, $\mbound$ starts to decrease exponentially, and $\rdetect$ coincides with the median value of $\rapo$ (inset in the top left panel of \autoref{fig:apos}).
Consequently, for the BRC to be detectable, $\rapo$ must be equal to $\rapop$, resulting in a $\theta_\mathrm{min} \rightarrow 90\fdg0$.
For recoil velocities $\vk\gtrsim 600\,\kmps$, the exponential increase in apocentre distance outpaces the linear increase in $\rdetect$, and so $\rapop>\rdetect$, resulting in a decreasing required angular offset $\theta_\mathrm{min}$ until the BRC mass is below the detection limits ($\vk\gtrsim1100\,\kmps$). 

We can estimate the fraction of galaxies that are likely to have a detectable recoiling SMBH given some recoil velocity by performing a third transformation sampling, now of the distribution of $\theta_\mathrm{min}$.
For each posterior draw $\tilde{v}_\mathrm{kick}$, there is an angular offset $\theta_\mathrm{min}$ which the SMBH must exceed to be detectable. 
Consider the galaxy centre as a circle in projection with area $A_\mathrm{occ}$, which occults two sides of a sphere with radius $\rapo$.
Then the fraction of surface which is not occulted becomes:
\begin{align*}
    A_\mathrm{not\;occ} &= 1 - 2A_\mathrm{occ} \\
    &= 1 - 2 \times \frac{2\pi(1-\cos\theta_\mathrm{min})}{4\pi} \\
    &= \cos\theta_\mathrm{min} \implies \in [0, 1].
\end{align*}
For each Monte Carlo sampled value of $\tilde{v}_\mathrm{kick}$ in \autoref{fig:apos}, we weight the value of $\tilde{v}_\mathrm{kick}$ by its correspondingly-sampled $\cos\theta_\mathrm{min}$ to obtain a distribution of kick velocities weighted by observational probability, shown as blue bars in the bottom panels in \autoref{fig:apos}.
A clear minimum is visible for $\vk\sim600\,\kmps$, corresponding to the extended $\theta_\mathrm{min}\sim90\fdg0$ region in the top right panel of \autoref{fig:apos}.
This is contrasted to the steadily-decreasing $\vk$ probability predicted by \citet{zlochower2015} for our system.
Taking the cumulative sum of relative kick velocities, we find that whilst $\sim45\%$ of recoiling SMBHs have $\vk\lesssim0.6v_\mathrm{esc}$, only $\sim20\%$ of all recoiling SMBHs would have a detectable BRC.

We stress that the fraction of detectable BRCs discussed here are for our particular initial conditions.
Whilst we do not test differing galaxy initial conditions directly, our expectation for steeper stellar density profiles would be an increase in the number of detectable BRCs at a given radius -- owing to the combination of more strongly-bound stellar particles to the SMBH at the time of binary coalescence, and the more rapid decrease in the galaxy luminosity with radius.
However, the deeper central potential of a galaxy with a steeper stellar density profile would likely reduce the apocentre a recoiling SMBH could achieve given some kick velocity $\vk$.
The complex interplay of these factors makes it difficult to assess the detectable fraction of BRCs when considering different density profiles, and is an avenue for future investigation.

\begin{table*}
    \centering
    \begin{tabular}{c|c|c|c|c|c}
    \hline\hline
        Instrument & Field of View (sq$\arcsec$) & Sampling ($\arcsec$/pixel) & Angular Resolution ($\arcsec$) & Slit Length ($\arcsec$) & Slit Width ($\arcsec$) \\
        \hline 
        MUSE & $7.42^2$ & 0.0250 & 0.200 & \dots & \dots \\
        HARMONI & $0.61^2$ & 0.0040  & 0.020 & \dots & \dots \\
        MICADO & $18.0^2$ & 0.0015 & 0.020 & 3.00 & 0.0163\\
        ERIS & $26.0^2$ ($0.84^2$) & 0.0130 (0.0250) & 0.100 & 12.0 & 0.068\\
        JWST & $204^2$ ($3.00^2$) & 0.1000 (0.1000) & 0.068 & 3.20 & 0.200\\
        \hline
    \end{tabular}
    \caption{
        Assumed instrument parameters used to construct the mock IFU and LSS images in \autoref{fig:ifu} and \autoref{fig:lss}.
        For ERIS and JWST, the numbers in brackets indicate the parameters for the IFU observations.
        For ground-based instruments, the quoted angular resolution exceeds design specifications to account for atmospheric distortion.
    }
    \label{tab:instruments}
\end{table*}

\section{Mock kinematic observations}\label{sec:kinematics}
\begin{figure*}[t]
    \centering
    \includegraphics[width=0.98\textwidth]{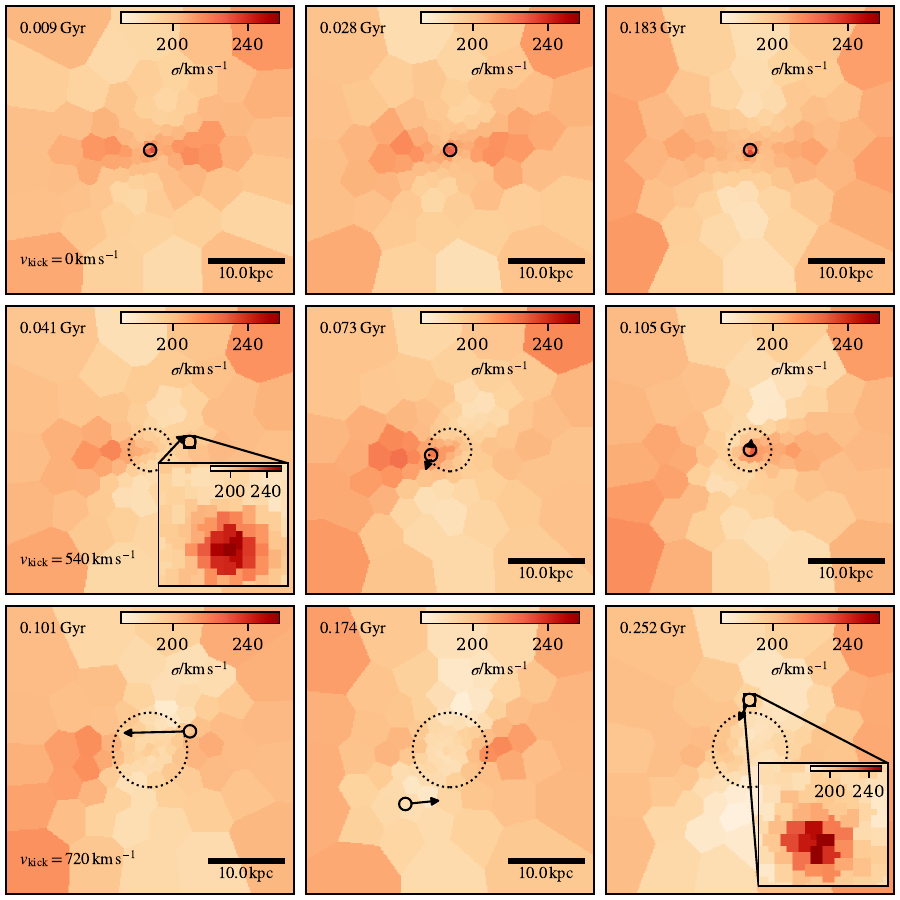}
    \caption{ 
        Mock MUSE IFU observations for three recoil velocities: $0\,\kmps$ (top row), $540\,\kmps$ (middle row), and $720\,\kmps$ (bottom row).
        Arrows indicate the instantaneous velocity vector of the SMBH (black circle), and the dotted circle the radius beyond which the BRC has a projected density above that of the galaxy.
        In the case of no SMBH recoil, there are symmetric `lobes' of increased $\sigma$, whereas for non-zero recoil velocity, asymmetry in the lobes are present, with the absence of a lobe indicating the approximate position of the SMBH.
        In the middle left and bottom right panels we show an inset mock HARMONI IFU map centred on the SMBH position in a $1.5\,\mathrm{kpc}$ aperture, highlighting the locally-enhanced velocity dispersion of the BRC.
    }
    \label{fig:ifu}
\end{figure*}
\begin{figure*}[t]
    \centering
    \includegraphics[width=0.98\textwidth]{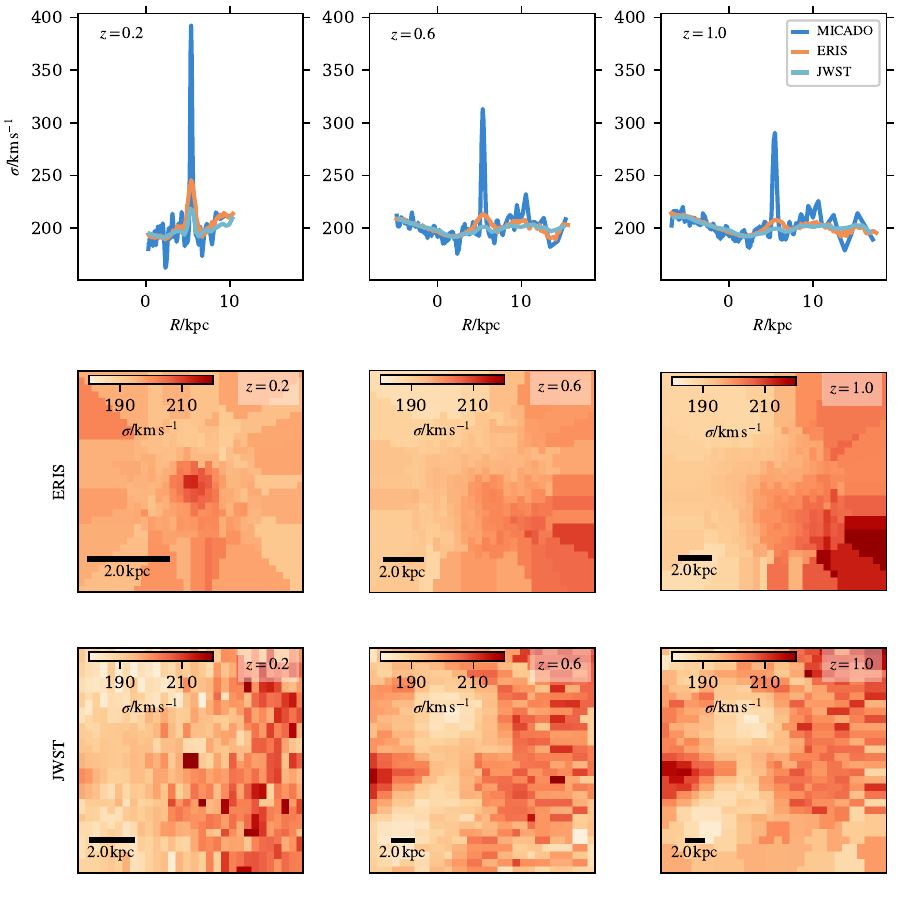}
    \caption{
        \textit{Top row}: Velocity dispersion profile (centred on the BRC at apocentre for the $\vk=540\,\kmps$ simulation) for three long-slit spectroscopy instruments: MICADO (a first-generation ELT instrument), ERIS, and JWST, with each column corresponding to a different observation redshift of the BRC.
        MICADO offers the best chance of detection at higher redshifts than either ERIS or JWST.
        \textit{Centre row}: Mock IFU maps for ERIS at different redshifts consistent with the top row.
        \textit{Bottom row}: Mock IFU maps for JWST at different redshifts consistent with the top row.
        For both rows, the IFU maps are centred on the SMBH.
        Consistent with long-slit spectroscopy observations, the BRC is visible in velocity dispersion only at the lowest redshift.
    }
    \label{fig:lss}
\end{figure*}
\subsection{Integral field unit observations with MUSE}\label{ssec:IFU_observations}
To confirm the presence of an SMBH within a BRC, it has been suggested \citep[e.g.][]{merritt2013} that the velocity dispersion of the BRC would be characteristically high.
However, taking stellar kinematic measurements of each individual cluster-like object in a photometric field would be observationally-expensive, and so we consider integral field unit (IFU) observations of the galaxy merger remnant to discern the imprint the recoiling SMBH leaves on the global stellar kinematics. 
The approach follows that done by \citet{naab2014} and \citet{rawlings2025} for the case where the SMBH is recoiling orthogonal to the LOS; we ignore the case of an SMBH recoiling parallel to the LOS, as the BRC would not be detected with photometric techniques.

The mock IFU maps are created assuming the specifications of the Multi Unit Spectroscopic Explorer (MUSE) narrow field mode (NFM) at the Very Large Telescope, which has a field of view of $7\farcs42$ and spatial sampling of $0\farcs025$ per pixel \citep[][ see \autoref{tab:instruments}]{bacon2010}, corresponding to a physical spatial extent of $\sim 50\,\mathrm{kpc}$ and pixel width of $0.171\,\mathrm{kpc}/\mathrm{pixel}$ at our fiducial redshift of $z=0.6$.
We take those stellar particles within a projected box of side length $40\,\mathrm{kpc}$, approximately the MUSE NFM field of view at our assumed redshift, and degrade the observation quality by generating 25 pseudo-particles for each stellar particle, spatially distributed as a normal distribution $\mathcal{N}(x_i, 1.372)$, where $x_i$ is the spatial position of stellar particle $i$ and the standard deviation is chosen as the angular resolution of the MUSE NFM instrument, both in kpc.
The pseudo-particles are then binned onto a rectangular grid with $40\,\mathrm{kpc}/(0.171\,\mathrm{kpc}/\mathrm{pixel})\sim230^2$ pixels, from which Voronoi bins are constructed using the method of \citet{cappellari2003}.
We require a Poisson S/N error of $N/\sqrt{N} = 1000$, resulting in $10^6$ pseudo-particles per Voronoi bin, to which we fit a \ordinal{4}-order Gauss-Hermite series (\autoref{sec:app_losvd}).

We show the mock IFU observations for the velocity dispersion for three recoil velocities each at three different times in \autoref{fig:ifu}.
In the first row, depicting the no-recoil case, there are two distinct, symmetric `lobes' of increased $\sigma$ that extend to $\sim 10\,\mathrm{kpc}$ either side of the SMBH position.
The lobes persist and do not vary with time since the SMBH binary coalescence.
We also create mock IFU observations in a different projection to confirm that the lobes are not an artefact of the particular viewing direction.
The lobe features arise during the dynamical friction phase of the SMBH-binary merger (i.e. prior to SMBH coalescence), also causing counter-rotating regions, as mass bound to the SMBH prior to merger is lost but retains an imprint of the galaxy merger orbit \citep{rantala2019,frigo2021}.
This feature, associated with central kinematically-decoupled regions, requires a major merger with $q\equiv M_{\star,1}/M_{\star,2} \gtrsim 1/3$ \citep{rantala2019}.

For non-zero recoil velocity, we observe systematic changes in the velocity dispersion arising from the recoiling SMBH, shown in the bottom two rows of \autoref{fig:ifu}.
The SMBH is brought to rest at apocentre, with successive apocentres decreasing due to dynamical friction with the surrounding stellar particles \citep{chandrasekhar1943}.

The particles in the vicinity of the SMBH at apocentre have their orbits disrupted, removing the right lobe in $\sigma$ at first apocentre (middle row, left in \autoref{fig:ifu}).

Conversely, if the SMBH apocentre is beyond the lobe (and hence the instantaneous velocity of the SMBH is greater than the stellar velocity dispersion as the SMBH passes through the lobe), it is not disrupted, as in the bottom left panel of \autoref{fig:ifu}.
Here, the lobe disruption is not as significant as in the $\vk=540\,\kmps$ case, where the SMBH apocentre coincided with the lobe position.

From the mock IFU observations, we suggest that asymmetric stellar velocity dispersion can be attributed to a recoiling SMBH, and can be used to also constrain where the SMBH \textit{recently} was.
How far the SMBH, and also its BRC, has travelled since the disruption of the $\sigma$ lobe will depend on the instantaneous SMBH velocity $v_\bullet$, which in turn depends on the original recoil velocity $\vk$ and the number of apocentres completed.

\subsection{Integral field unit observations with HARMONI}\label{ssec:harmoni}
As a proof-of-concept, we also create two mock IFU observations using the specifications for the High Angular Resolution Monolithic Optical and Near-infrared Integral field spectrograph (HARMONI), a first-generation Extremely Large Telescope instrument \citep{thatte2010,thatte2016}.
Compared to MUSE, MICADO will boast $0\farcs01$ spatial sampling, corresponding to a pixel width of $0.07\,\mathrm{kpc}/\mathrm{pixel}$ at the fiducial redshift $z=0.6$.
We take a $1.5\,\mathrm{kpc}$ window centered on the SMBH location at an arbitrary time for the $\vk=540\,\kmps$ simulation (middle left panel of Figure \ref{fig:ifu}) and the $\vk=720\,\kmps$ simulation (bottom right panel in Figure \ref{fig:ifu}).
We again generate 25 pseudo-particles per stellar particle to create the mock IFU images, shown as the inset in the middle left and bottom right panels of \autoref{fig:ifu}.
We see a clearly elevated $\sigma$ coinciding with the BRC location, with values exceeding $240\,\kmps$ in both instances.
An offset peak in velocity dispersion of the order of several hundreds of kilometres per second would indicate an exciting candidate for a recoiling SMBH, making future instruments such as HARMONI ideal for such detections.

\subsection{Long-slit spectroscopy}
We also create mock observations of the BRC using long-slit spectroscopy (LSS).
We consider three instruments: the upcoming Multi-AO Imaging Camera for Deep Observations (MICADO), another first-generation Extremely Large Telescope instrument \citep{sturm2024}, as well as the Enhanced Resolution Imager and Spectrograph \cite[ERIS, mounted at the VLT, e.g.][]{davies2023} and the recently-launched JWST NIRSPEC \citep[e.g.][]{rieke2023}, both currently available for observing.

We simulate the velocity dispersion profile through a narrow slit centred on the BRC location for the $\vk=540\,\kmps$ case at apocentre (i.e. the same snapshot as the centre left panel of \autoref{fig:ifu}) by generating 100 pseudo-particles for each stellar particle located (in projection) within the slit distributed as $\mathcal{N}(x_i, \tau)$  (where $\tau$ is the angular resolution of the instrument), binning these pseudo-particles into equal-mass bins of 100 per bin.
If the spatial separation between a bin $i$ and its neighbour $i+1$ is less than the instrument spatial sampling (\autoref{tab:instruments}), we join those bins.
In this way, we ensure that we have a spatial sampling set by the design specifications of the particular instrument.

The velocity dispersion profile is determined for each of MICADO, ERIS, and JWST for each of $z=\{0.2, 0.6, 1.0\}$, and is shown in the top row of \autoref{fig:lss}.
At $z=0.2$, the BRC is seen as a clear peak in the profile for each instrument, located at $\sim5\,\mathrm{kpc}$.
The high spatial sampling of MICADO is best able to resolve the BRC, with a $\sigma\simeq 400\,\kmps$ peak.
A $\sigma\simeq 250\,\kmps$ peak is resolved with both ERIS and JWST.
At higher redshifts, only MICADO is able to resolve the peak in the velocity dispersion, with the dispersion peak decreasing with redshift due to the increasing angular scale. 
ERIS and JWST are barely able to resolve the BRC at $z=0.6$, and not at all at $z=1.0$, with the velocity dispersion profile appearing relatively flat (including at the centre of the galaxy).
The mock LSS observations clearly highlight MICADO as a powerful tool in resolving the enhanced velocity dispersion of a BRC.

\subsection{Integral field unit observations with ERIS and JWST}
In addition to long-slit spectroscopy, ERIS and JWST also possess the capability to perform IFU observations. 
We create mock IFU maps for each instrument following the approach in \autoref{ssec:IFU_observations}, and shown in the middle and bottom rows of \autoref{fig:lss}.

In agreement with the LSS results, the BRC is resolved as a localised peak in velocity dispersion at a low redshift of $z=0.2$, whereas at higher redshifts, the BRC is not visible.
Hence, whilst detecting nearby BRCs might be possible with both instruments, a clear IFU detection in $\sigma$ will require the upcoming capabilities of HARMONI.

\subsection{Distinguishing a BRC from other objects}\label{ssec:obscomp}
\begin{figure*}[t]
    \centering
    \includegraphics[width=0.98\textwidth]{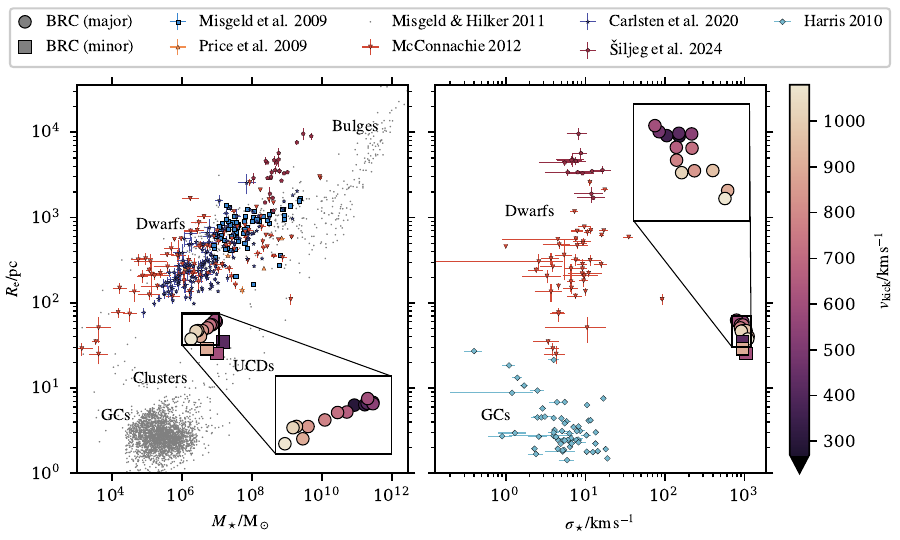}
    \caption{
        Comparison of the BRCs (coloured by $\vk$) to known faint objects: dwarf galaxies, ultra-compact dwarfs (UCDs), ultra-diffuse galaxies, and Milky Way globular clusters.
        Whilst the BRC has a similar mass and effective radius to some faint objects (left panel), it is clearly distinct from these objects in its velocity dispersion signal (right panel).
        This is the case for BRCs resulting from both major (circle points) and minor (square points) mergers.
        For visual clarity, we only show the major merger simulations in the inset panels.
    }
    \label{fig:obs}
\end{figure*}
In practice, photometric observations of galaxy merger remnants will contain contamination from other `intruding' stellar systems, globular clusters, satellites, or chance alignment with distant background galaxies, potentially complicating the unambiguous detection of a BRC about a recoiling SMBH.
To compare against observations of faint objects, we compute at apocentre $\mbound$ (consistent with \autoref{fig:bound}), the effective radius $\BRCR$, and the stellar velocity dispersion $\BRCV$ of the BRC within the BRC effective radius $\BRCR$.
The BRC velocity dispersion we compute by bootstrapping the velocities of the member stellar particles to reduce random noise, with $10^4$ samples.
\nocite{misgeld2009,price2009,harris2010,misgeld2011,mcconnachie2012,carlsten2020,siljeg2024} 

In the left panel of \autoref{fig:obs}, increasing $\vk$ corresponds to a constant decrease of the effective radius of the BRC.
A decrease in $\mbound$ with increasing $\vk$ is also visible -- the resulting effective stellar surface mass densities of the BRCs are in the range of $\Sigma_\mathrm{eff}\sim 50 - 100\,\Msun\,\mathrm{pc}^{-2}$.
Both trends are explained due to a higher $\vk$ requiring stellar particles to be more strongly-bound to the BRC than a low $\vk$.
In the $M_\star$-$\Reff$ plane, the BRCs mostly resemble an intermediary between dwarf systems and ultra-compact dwarfs (UCDs), that are slightly more compact than the former for a given stellar mass.

More effective in separating the BRCs from other faint objects is the stellar velocity dispersion $\BRCV$: in general, the BRCs have a high velocity dispersion of the order 100 times greater than other compact systems.

In line with theoretical expectation \citep[e.g.][]{merritt2009}, the velocity dispersions in \autoref{fig:obs} grow with increasing $\vk$, again owing to higher $\vk$ SMBHs requiring more tightly bound stellar particles to constitute the BRC.

One observed object has a velocity dispersion $\sim 100\,\kmps$ and a relatively compact effective radius: this is the dwarf system M32, which has evidence for a central SMBH \citep{vandermarel1997,mcconnachie2012}, however is unlikely to be a BRC hosting a recoiling SMBH, given the very large stellar mass of the system ($\sim10$ times that of our simulated BRCs).

To our knowledge, there is currently no faint object with a velocity dispersion consistent with our findings of the simulated BRCs, highlighting the need for targeted kinematic campaigns to uncover these objects.

\section{Minor mergers}
Our investigation has thus far centred around the major merger of two equal-mass galaxies, and hence two equal-mass SMBHs.
This introduces a bias towards larger recoil velocities that take the SMBH further from the galaxy centre than what might be expected for an unequal-mass merger.
To investigate the robustness of our analysis, we consider a minor merger of mass ratio $1:5$ (where the less massive galaxy has its stellar mass reduced by a factor of 5) motivated by \citet{oser2012}, who found this to be the dominant mass-weighted mass ratio for mergers involving massive galaxies.
The SMBH mass and DM mass are adjusted with the stellar mass, resulting in $M_\bullet=5.8\times10^8\,\Msun$ and $M_\mathrm{DM}=5.0\times10^{12}\,\Msun$.
We test recoil velocities $\vk=[0, 420, 600, 900]\,\kmps$.
We find that the BRCs in these simulations have similar masses and velocity dispersions to the BRCs from equal-mass mergers, but are slightly more compact in effective radius (shown as the square points in \autoref{fig:obs}).

We find that unequal-mass mergers can produce asymmetric velocity dispersions in the mock IFU images, potentially presenting a challenge for identifying a BRC.
However, any potential degeneracy between identifying a BRC and the presence of an unequal-mass merger can be broken by considering the first order asymmetric coefficient $h_3$ contribution to the LOSVD.
In the instance of a recoiling BRC, there is no strong spatial correlation in $h_3$ across Voronoi cells in the mock IFU images -- the $\sigma$ lobes are altered by the perturbation of a massive point-mass potential. 
Conversely, strong spatial correlation in $h_3$ is visible for unequal-mass mergers, owing to the large amount of stellar mass that alters the stellar potential as the merger remnant settles, essentially contributing a kinematically-decoupled component to the LOSVD.
We show example maps in \autoref{fig:ifu_comp} in \autoref{sec:app_comp}.

This does however highlight the need to consider multiple pieces of evidence for a recoiling BRC before claiming a detection, which we discuss briefly now in the context of a suggested workflow for identification.

\section{Discussion and Summary} \label{sec:discussion}
\begin{figure}[t]
    \centering
    \includegraphics[width=0.48\textwidth]{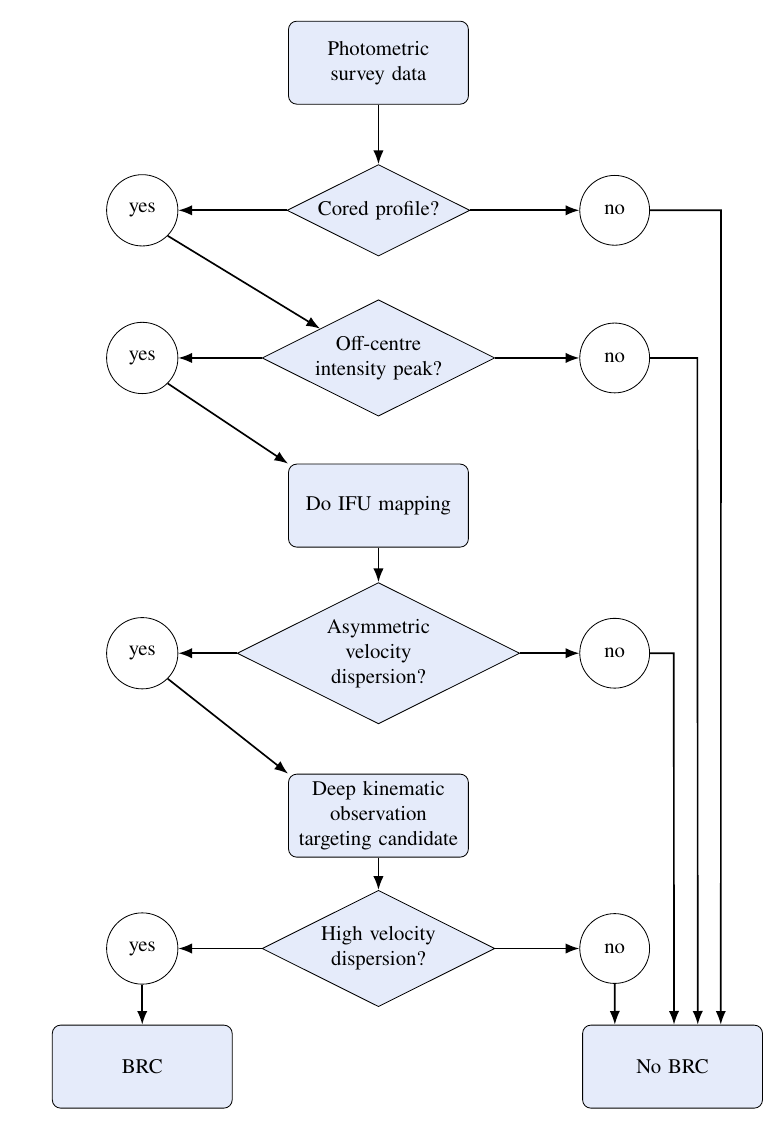}
    \caption{
        Suggested workflow to identify a BRC, starting from photometric data. 
        At each decision point (diamonds), the sample of candidate-BRCs can be reduced, leaving a smaller sample for the most time-expensive observations.
    }
    \label{fig:flow}
\end{figure}
Our results present a workflow to identify SMBH-hosting BRCs, and is summarised in \autoref{fig:flow}.
Candidate BRCs may be found in large-scale photometric surveys, such as those from Euclid and the upcoming Roman Telescope, as significantly-brighter objects than their local surrounds (\autoref{fig:projdens}), drastically reducing the number of targets required for follow-up kinematic observations.
As highlighted in \autoref{fig:obs}, the clearest signature of a BRC is an abnormally-high stellar velocity dispersion.
Kinematic measurements of the candidate BRCs can be performed, or if the number of BRC candidates is prohibitively large, an additional cut can be achieved by identifying asymmetry in galactic stellar velocity dispersion from an IFU survey (\autoref{fig:ifu}) before performing the BRC candidate-specific kinematic observations.

From \autoref{eq:bound}, a more massive SMBH is more likely to host a detectable BRC than a less massive SMBH.
Consequently, the prime targets for investigation are those galaxies which are expected to host the most massive SMBHs, namely cored galaxies \citep[e.g.][]{thomas2016}.
A large stellar core in an ETG is most readily attributed to SMBH binary scouring \citep[e.g.][]{Thomas2014}: observing a core indicates that a SMBH binary was likely present in the galaxy at some point, and hence a recoil candidate is plausible.
Assuming a stellar core radius of the order $1\,\mathrm{kpc}$, corresponding to an angular size of $\sim0\farcs06$ at $z=0.6$, massive cored ETGs below this redshift are an ideal first place to look for BRCs with current instrument capability.

A key result of our study is the dependence of BRC observability on recoil velocity (\autoref{fig:apos}).
We found a recoil velocity of $\vk\sim 540\,\kmps$ to represent a particularly challenging velocity to detect: mass of the BRC is less than that for lower recoil velocities, and the low SMBH $\rapo$ reduces the contrast of the BRC compared to the host galaxy.
We expect that for different initial conditions, the most challenging BRC to observe would correspond to $\rdetect\simeq \Reff$, however this would most likely be in the vicinity of $\vk\sim600\,\kmps$ given most massive ETGs have $\Reff\sim6.0\,\mathrm{kpc}$ \citep[e.g.][]{kelvin2012}, similar to our models with $\Reff\sim5.6\,\mathrm{kpc}$.
Comparatively, the most likely BRCs to be detected are those which have $\vk>\sigcore$ and $\rdetect < \Reff$: these kicks are more frequent than $\vk \simeq 0.5v_\mathrm{esc}$, and are accompanied by a BRC massive enough for detection.

In this work, we have focused primarily on SMBH recoil orthogonal to the LOS axis.
The majority (55\%) of kicked SMBHs recoil with an angle to the LOS axis that results in $\rapo$ less than the detection threshold distance $\rdetect$: these BRCs potentially have a velocity component parallel to the LOS axis of the order $\sim\vk$, and might be detectable as an asymmetric tail in the LOS velocity distribution extending to $v_\mathrm{LOS}\sim\vk$.
This signature would be clearly attributable to an ejected SMBH: such a mean BRC velocity is unobtainable with regular stellar dynamics.
Such a discovery is however unlikely as the instantaneous velocity of a recoiling SMBH approaches zero at apocentre, making the time for which the BRC has $v\gg \sigma$ increasingly short and hence unlikely to be detected. 

This work has focused on detections of recoiling SMBHs by way of their enshrouding stellar mass, however the techniques presented here may also pave an avenue for detecting wandering black holes that have been ejected from lower mass systems, either dynamically or due to GW-induced recoil \citep[e.g.][]{rizzuto2022,rantala2024b,rantala2025}.
Wandering black holes formed in nuclear stellar clusters are also expected to posess an entourage of bound stellar material, however as these black holes are generally of lower mass than those considered in this work, one would expect their stellar velocity dispersions to be lower, of the order a few tens to a hundred $\kmps$ \citep[e.g.][]{stone2017}.

To give an order-of-magnitude estimate for the number of detectable BRC-hosting systems, let us consider the number of major mergers ($1/3\lesssim \xi \leq 1$, where $\xi\equiv M_2/M_1$) a halo of mass $M_\mathrm{DM}\sim 5\times10^{13}\,\Msun$ (i.e. the simulated merger remnant in this work) would have between $0\leq z \leq 1$.
Let us further assume that each halo hosts a massive galaxy with a SMBH, and that each halo merger results in the coalescence of the SMBHs before $z=0$.
The expected number of mergers per halo satisfying our constraints is given by:
\begin{equation}
    N_\mathrm{m}(\xi_0, M_0, z_0, z) = \int_{z_0}^{z} \dd{z'} \int_{\xi_0}^1 \dd{\xi'} \dv{N_\mathrm{m}}{\xi'\dd{z'}}\left( M(z'),\xi',z' \right)
\end{equation}
where $\dv{N_\mathrm{m}}{\xi\dd{z}}(M,\xi,z)$ is fit to the collisionless Millennium and Millennium II simulations in \citet{fakhouri2010}.
Multiplying this result with the cumulative halo mass function for $M_\mathrm{DM}\gtrsim 5\times10^{13}\,\Msun$ \citep[e.g.][]{rodriguezpuebla2016} results in a $\sim 2\times10^{-4}\,\mathrm{Mpc}^{-3}$ major merger density.
Hence, out to a redshift $z=1$ where the comoving volume is $\sim1.7\times10^{11}\,\mathrm{Mpc}^3$, there are $\sim4\times10^7$ mergers satisfying our constraints, and a corresponding number of recoiling SMBHs.
Repeating the exercise for $z=0.6$ results in a value of $\sim 7\times10^6$ recoiling SMBHs.
We estimated $\sim20\%$ of recoiling SMBHs would \textit{at some point} in their excursion from the galaxy centre be observable (\autoref{ssec:offaxis}); this translates to $\sim8\times10^6$ BRCs out to $z=1$.
Estimates for \textit{how long} such BRCs would be observable is significantly more complicated, as torquing of the SMBH away from the initial recoiling axis can increase the amount of time the SMBH is below the detection limit $\rdetect$ in projection.
Clearly, the largest window of observability occurs for purely radial motion of the settling SMBH, and is thus most probable up until first apocentre, which we found to have a median value of $t_\mathrm{apo}\sim30\,\mathrm{Myr}$.
Thus, when considering the observable duration of a BRC, we may roughly estimate the number of detectable BRCs to be:
\begin{equation}\label{eq:numBRCs}
    N_\mathrm{detections} \approx N_\mathrm{m}(z) f_\mathrm{obs} \frac{t_\mathrm{apo}}{t_\mathrm{lookback},(z)}
\end{equation}
which for $z=1$ gives $N_\mathrm{detections}\sim3\times10^4$, and some $8\times10^3$ for $z=0.6$, assuming $f_\mathrm{obs}\simeq 0.2$. 
Consequently, whilst not common (particularly compared to the other faint objects shown in \autoref{fig:obs}), we expect there to be a small population of BRCs waiting to be uncovered by large-scale extragalactic surveys, and then confirmed with kinematic observations.

To provide robust estimates for the total duration for which a BRC would be observable, thus refining the estimate given by \autoref{eq:numBRCs}, one would need to extend the study performed here to include a variety of galaxy triaxialities (our merger remnant has a triaxiality of $T=\left( a^2 - b^2 \right) / \left( a^2 - c^2 \right) \simeq 0.82$), obtained through cosmological simulations where the impact of the surrounding environment could be quantified.
Additionally, one might consider the effect stochastic properties of the SMBH binary prior to coalescence might have on the BRC, and hence its detectability.
For example, a low SMBH binary eccentricity arising from small phase space perturbations \citep[e.g.][]{rawlings2023} would increase the overall time spent in the binary scouring phase before GW-driven coalescence, lowering the central density \citep{rantala2024} and potentially also the amount of stellar mass available to form the BRC. 

Despite the modelling assumptions, our results provide an important step in the hunt for SMBH recoil. 
Advances in instrumentation, the capabilities offered by ongoing and upcoming large-scale photometric surveys, coupled with ever-improving kinematic observations, work in unison to identify SMBH recoil in action. 
Comparing our simulations to observations of faint objects in the Universe, it appears that candidates for SMBH-hosting BRCs presently eludes us.

\section{Acknowledgments}
We thank the anonymous referee for their comments which helped improve the manuscript.
We also thank Till Sawala and Maximilian Fabricius for useful discussions.
A.R. acknowledges the support by the University of Helsinki Research Foundation.
A.R. and P.H.J. acknowledge the support by the European Research Council via ERC Consolidator Grant KETJU (no. 818930).
P.H.J also acknowledges the support of the Research Council of Finland grant 339127.
T.N. acknowledges support from the Deutsche Forschungsgemeinschaft (DFG, German Research Foundation) under Germany’s Excellence Strategy - EXC-2094 - 390783311 from the DFG Cluster of Excellence ``ORIGINS''.
This research has made use of the SVO Filter Profile Service ``Carlos Rodrigo'', funded by MCIN/AEI/10.13039/501100011033/ through grant PID2023-146210NB-I00.
We list here the roles and contributions of the authors according to the Contributor Roles Taxonomy (\href{https://credit.niso.org}{CRediT}). 
\textbf{AR}: conceptualisation, methodology, formal analysis, investigation, data curation, writing: original.
\textbf{PHJ}: supervision, writing: original, funding acquisition.
\textbf{TN}: conceptualisation, investigation, writing: original.
\textbf{AR}: writing: review.
\textbf{JT}: writing: review.
\textbf{BN}: writing: review.

%

\software{
\ketju{} \citep{mannerkoski2023,rantala2017},
\gadget{} \citep{springel2021},
NumPy \citep{harris2020},
SciPy \citep{virtanen2020},
Matplotlib \citep{hunter2007},
pygad \citep{rottgers2020},
\textsc{Stan} \citep{standevelopmentteam2018},
CmdStanPy \citep{standevelopmentteam2018},
Arviz \citep{kumar2019},
Synthesizer \citep{vijayan2021}.
}



\appendix

\section{Defining prominence with a Gaussian filter}\label{sec:app_prom}
To identify the BRC in mock photometric observations, we use a low-pass Gaussian filter with a kernel size parametrised by $\tau=0.5$ pixels.
The filter is defined by convolving a Gaussian kernel at pixels $(u,v)$, $G_\tau(u,v)$, with the image $F$ within $4\tau\rightarrow[-k,...,+k]$ of the central pixel located at $(i,j)$, and updating the central pixel value with this weighted average:
\begin{equation}
    K(i, j) = \sum_{u=-k}^{+k} \sum_{v=-k}^{+k} \frac{1}{2\pi\tau^2} \exp\left( -\frac{u^2 + v^2}{2\tau^2}\right) F(i-u, j-v).
\end{equation}

We then standardise the pixel values $K(i,j)$ using Gaussian statistics:
\begin{equation}
    \hat{K}(i,j) = -\frac{K(i,j)-E[K(i,j)]}{\sqrt{\mathrm{var}[K(i,j)]}},
\end{equation}
where the leading negative sign indicates lower-than-expected magnitude pixels (i.e. pixels with higher flux) have a positive prominence, and the units are that of standard deviation.
To define a detection of the BRC, we then perform a standard one-tailed hypothesis test to determine if $\hat{K}(i,j)$ is significantly larger than 0 (and hence significantly brighter than its surrounds).
We require for our detection that $\hat{K}(i,j) > 4$, meaning that we can reject the null hypothesis, that $\hat{K}(i,j)\leq0$, with a $p$-value of $p\sim10^{-5}$.
This means that a value of $\hat{K}(i,j)>4$ would occur once in every $10^5$ instances by chance alone.

\section{Line of sight velocity distribution}\label{sec:app_losvd}
To generate the mock IFU images, pseudo-particles are divided using a Voronoi tesselation.
The LOS velocity distribution of pseudo-particles within each Voronoi bin are then fit using a Gauss-Hermite series (after fitting for the mean velocity $V$ and velocity dispersion $\sigma$) defined \citep{vandermarel1993}:
\begin{equation}
    \mathscr{L}(w) = \frac{1}{\sqrt{2\pi}\sigma} e^{-w^2/2} \left\{ 1 + \sum_{j=3}^n h_j H_j(w) \right\},
\end{equation}
where $w \equiv (v_\mathrm{LOS} - V)/\sigma$, and the Hermite polynomials $H_j(w)$ are:
\begin{equation}
    H_j(w) = (-1)^j e^{w^2} \dnv{j}{e^{-w^2}}{w}.
\end{equation}
Whilst we use a \ordinal{4} order fit in this work, we tested both $2^\mathrm{nd}$ and \ordinal{6} order fits, however find the former to best match the LOSVD profile.

\section{IFU comparison between major and minor mergers}\label{sec:app_comp}
\begin{figure}[t]
    \centering
    \includegraphics[width=0.68\textwidth]{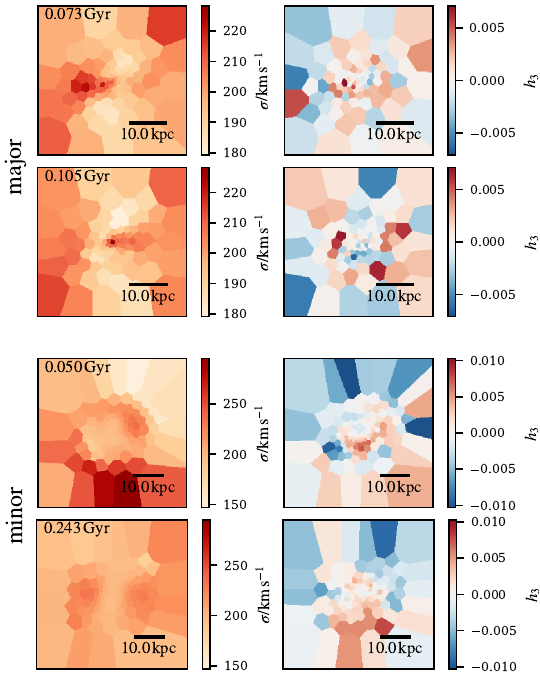}
    \caption{
        Comparison between major mergers (top two rows) and minor mergers (bottom two rows) in velocity dispersion $\sigma$ (left column) and first order asymmetric deviation $h_3$ (right column).
        The times are normalised such that $t=0$ corresponds to the time of SMBH binary coalescence.
        The major merger has a recoiling SMBH with $\vk=540\,\kmps$, and is the same IFU maps as in the middle centre and middle right panels of \autoref{fig:ifu}.
        Clear asymmetric lobes in velocity dispersion are present, but no strong spatial correlation between Voronoi pixels in $h_3$ are visible.
        The minor merger has no recoiling SMBH, with the seen velocity dispersion asymmetry at $t=0.05,\mathrm{Gyr}$ arising from the asymmetry in the merger.
        At this moment there is strong spatial correlation in $h_3$ at small radii, with evidence for two kinematically-distinct regions.
        As the merger remnant settles, there is a symmetric velocity dispersion, reminiscent of the no-recoil equal-mass merger in \autoref{fig:ifu}. 
        The $h_3$ map has reduced spatial correlation.
    }
    \label{fig:ifu_comp}
\end{figure}

    We show the mock IFU maps for the $\vk=540\,\kmps$, equal-mass merger remnant at two moments following SMBH coalescence, and similar maps for the no recoil unequal-mass merger, in \autoref{fig:ifu_comp}.
    We see that whilst unequal-mass mergers can produce asymmetric velocity dispersion features, as do recoiling SMBHS, the former also produces coherent features in the first order asymmetric deviation coefficient $h_3$, which a recoiling SMBH does not.


\bibliography{ref}{}
\bibliographystyle{aasjournal}



\end{document}